# On Modeling Dependency between MapReduce Configuration Parameters and Total Execution Time

Nikzad Babaii Rizvandi[1,2], Albert Y. Zomaya[1], Ali Javadzadeh Boloori[1,2], Javid Taheri[1],

[1] Center for Distributed and High Performance Computing, School of Information Technologies,
University of Sydney, Sydney, Australia

[2] Networked Systems Theme, National ICT Australia (NICTA), Australian Technology Park, Sydney, Australia

nbab6945 | ajav4801@uni.sydney.edu.au
albert.zomaya | javid.taheri@sydney.edu.au

*Abstract*—In this paper, we propose an analytical method to model the dependency between configuration parameters and total execution time of Map-Reduce applications. Our approach has three key phases: profiling, modeling, and prediction. In profiling, an application is run several times with different sets of MapReduce configuration parameters to profile the execution time of the application on a given platform. Then in modeling, the relation between these parameters and total execution time is modeled by multivariate linear regression. Among the possible configuration parameters, two main parameters have been used in this study: the number of Mappers, and the number of Reducers. For evaluation, two standard applications (WordCount, and Exim Mainlog parsing) are utilized to evaluate our technique on a 4-node MapReduce platform.

## I. INTRODUCTION

Recently, businesses have started using Map-Reduce as a popular computation framework for processing large-scaled data in both public and private clouds; e.g., many Internet endeavors are already deploying Map-Reduce platforms to analyze their core businesses by mining their produced data. Therefore, there is a significant benefit to application developers in understanding performance trade-offs in Map-Reduce-style computations in order to better utilize their computational resources [3-4].

Map-Reduce users typically run a few number of applications for a long time. For example, Facebook, which is based on Hadoop (Apache implementation of Map-Reduce in Java), uses Map-Reduce to read its daily produced log files and filter database information depending on the incoming queries. Such applications are repeated million times per day in Facebook. Another example is Yahoo where around 80-90% of their jobs is based on Hadoop[5]. The typical applications here are searching among large quantities of data, indexing the documents and returning appropriate information to incoming queries. Similar to Facebook, these applications are run million times per day for different purposes.

One of the existing challenges in this parallel processing context is to estimate the total execution time of an application. This problem becomes appealing as there exists a high correlation between tweaking/tuning MapReduce configuration parameters [6], such as number of Mappers or Reducers, and proper execution time of an application on MapReduce.

The proposed approach in this paper is a preliminary step towards modeling the relation between the MapReduce configuration parameters and the execution time of an application. Two major MapReduce configuration parameters investigated in this paper are the number of Mappers, and the number of Reducers. For a given MapReduce platform, applications run iteratively with different values of these parameters and the execution time of each run is extracted. Subsequently, a linear model is constructed by applying multivariate linear regression on the set of these configuration parameters values (as input) and execution time as output.

Obviously, the proposed modeling technique can be extended for other configuration parameters or used for modeling other resources such as storage, network bandwidth and memory. Although our modeling technique can be applied to other applications on different platforms, two issues should be concerned: firstly, the obtained model of an application on a specific platform may not be used for predicting the same application on another platform and secondly, the modeling of an application on a platform is not applicable to predicting other applications on the same platform

To demonstrate our approach, section II highlights the related works in this area. Section III provides the problem we focused in this paper. Section IV explains our analytical approach to profile, model and predict MapReduce applications' execution times followed by experimental results and conclusion in sections V and VI.

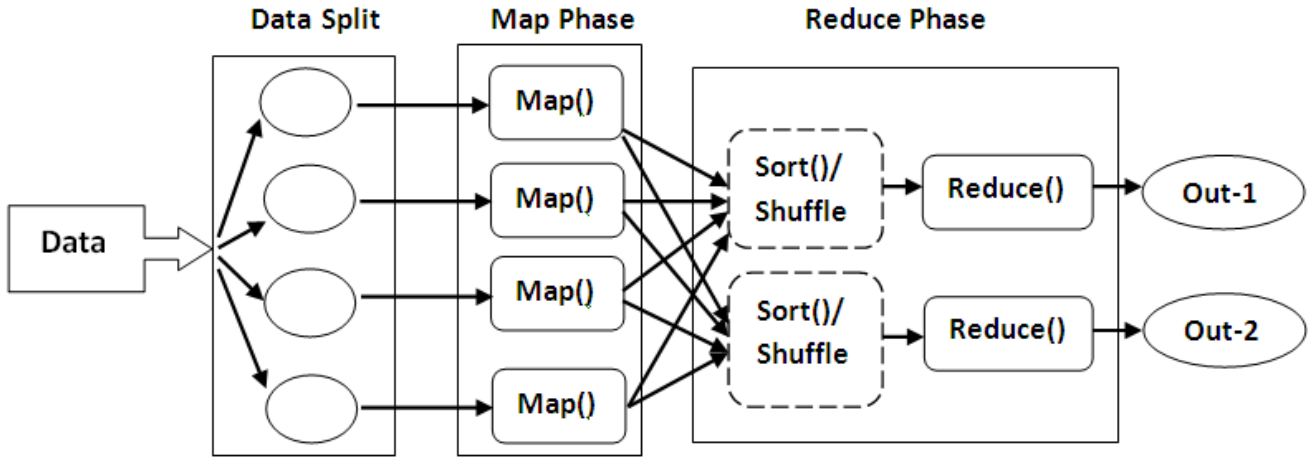

*Figure 1.MapReduce workflow [1-2]*

## II. RELATED WORKS

Early works on analyzing/improving Map-Reduce performance started almost since 2005; such as an approach by Zaharia et al [7] addressed problem of improving the performance of Hadoop for heterogeneous environments. Their approach was based on the critical assumption in Hadoop that works on homogeneous cluster nodes where tasks progress linearly. Hadoop utilizes these assumptions to efficiently schedule tasks and (re)execute the stragglers. Their work introduced a new scheduling policy to overcome these assumptions. Besides their work, there are many other approaches to enhance or analysis the performance of different parts of Map-Reduce frameworks, particularly in scheduling [8], energy efficiency [3, 9-15] and workload optimization[16]. A statistics-driven workload modeling was introduced in [10] to effectively evaluate design decisions in scaling, configuration and scheduling. The framework in this work was utilized to make appropriate suggestions to improve the energy efficiency of Map-Reduce[17].

A modeling method was proposed in [18] for modeling the total execution time of only Hive queries, a higher level software for database interaction written for Hadoop. It used Kernel Canonical Correlati[19-20]on Analysis to obtain the correlation between the performance feature vectors extracted from Map-Reduce job logs, and map time, reduce time, and total execution time. These features were acknowledged as critical characteristics for establishing any scheduling decisions. A basic model for Map-Reduce computation utilizations was presented in [21-22]. Here, at first, the map and reduce phases were modeled using dynamic linear programming independently; then, these phases were combined to build a global optimal strategy for Map-Reduce scheduling and resource allocation. . In [2, 23], an approach has been described to compare the CPU time pattern of a new unknown application with the CPU time pattern of several applications in database in order to find the most similar patterns. With regard to pattern matching concepts, it was concluded that if two applications have similar CPU patterns for several experiments with different values of parameters, it is much likely that the optimal values of configuration parameters of one application is applied to optimally run another application. In [24], we have applied linear regression to model the total number of CPU tick clocks an application needs to execute and four MapReduce configuration parameters. These configuration parameters are: number of Mappers, number of Reducers, size of file system and size of input file. In this paper, we almost follow the same concept by this paper but for modeling execution time of an application instead of the total CPU tick clocks.

## III. PROBLEM STATEMENT

Map-Reduce, introduced by Google in 2004 [25], is a framework for processing large quantities of data on distributed systems. The computation of this framework has two major phases: Map and Reduce as shown in figure 1.
In the Map phase, after copying the input file to the Map-Reduce file system and splitting the file into smaller files, data inside the split files are converted into $< key, value >$ format (e.g. $key$ can be a line number and $value$ can be a word in an essay). These $< key, value >$ pairs are entered to the Mappers and the first part of processing are applied on them. In fact, as Mappers in such framework are designed independently, Map-Reduce applications are always naturally ready for parallelization. This parallelization, however, can be bounded sometimes because of other issues such as the nature of the data source and/or the numbers of CPUs have access to the data.

In the Reduce phase, after finishing the Map phase, a network intensive job starts to collect intermediate produced $<key, value>$ pairs by Mappers to Reducers. Here, depending on the Map-Reduce configuration, a sort/shuffle stage may also be applied to expedite the whole process. Afterwards, map operations with the same intermediate $key$ will be presented to the same Reducers. The result is concurrently produced and written in output files (typically one output file) in the file system.

The process of converting an algorithm into independent Mappers and Reducers causes Map-Reduce to be inefficient for algorithms with sequential nature. In fact, Map-Reduce is designed for computing on significantly large quantities of data instead of making complicated computation on a small amount of data [26]. Due to its simple structure, Map-Reduce is suffering from serious issues, particularly in scheduling, energy efficiency and resource allocation.

In distributed computing systems, Map-Reduce has been known as a large-scale data processing technique [6, 26-27] indicating that execution time is one of the most important part of running an application on Map-Reduce. Therefore, reducing the total execution time of an application becomes important for customers to hire enough resources from cloud providers as well as for cloud providers to schedule incoming jobs properly.

Among different parameters influencing the total execution time of a running application on MapReduce cluster, in this paper we will study the influence of configuration parameters, specifically the values for number of Mappers and Reducers. Generally, problems regarding to the dependency between configuration parameters and performance outputs of applications in a framework, here MapReduce, can be grouped into four categories:

- Which configuration parameters are involved in effecting the output (such as amount of CPU utilization, Energy consumption, or execution time) for CPU-intensive applications? For MapReduce, the list of parameters can be found in [28].

- Among these parameters, which parameters are more important than the others? In other words, what are the most effective parameters influencing the output? This question can be answered by P-value, linear and non-linear correlation analysis for several experiments of application.

- How to mathematically model the relation between the effective parameters and output? In this paper, we will give an answer to this question by modeling the relation between execution time and two main parameters in MapReduce. Moreover, an approach has been proposed in [24] to model the relation between four configuration parameters and total CPU tick clocks for MapReduce applications. To be more precise, it is better to use non-linear modeling techniques like neural network.

- If a new unknown application arrives, how to find the closest known application in the database to it? This type of problem has been described in [23] for MapReduce applications.

The answers to the above categories can be applied to efficient managing of incoming jobs to a cluster/cloud by making scheduler smarter.

## IV. MODEL GENERATION

### A. Profiling

For each application, we generate several experiments with different sets of the number of Mappers/Reducers values on a given platform. After running each experiment, the total execution time of the application is extracted for future use as training data for the model Because of the temporal changes, it is expected that several runs of an experiment with the same configuration parameters may result in slightly different total execution time. Therefore, utilizing a mechanism to prune unsuitable data from the training dataset will improve the modeling accuracy. In [29], Robust Stepwise Linear Regression was used as a post processing stage to refine the outcome of the model by giving weights to data points with high error. In this study, we run an experiment five times and then the mean of these total execution time values is chosen as the total execution time of this experiment.

### B. Model generation

This section describes how to model the relation between the configuration parameters and total execution time of an application in MapReduce. The problem of modeling based on multivariate linear regression involves choosing the suitable coefficients of the modeling such that the model's response well approximates the real system response.

Consider the linear algebraic equations for M number of experiments of an application for $N$ effective configuration parameters ($M \gg N$) [24, 30-32]:

$$\begin{cases} T^{(1)} = \alpha_0 + \alpha_{11}p_1^{(1)} + \alpha_{12}\left(p_1^{(1)}\right)^2 + \alpha_{13}\left(p_1^{(1)}\right)^3 + \\ \quad \ldots + \alpha_{N1}p_N^{(1)} + \alpha_{N2}\left(p_N^{(1)}\right)^2 + \alpha_{N3}\left(p_N^{(1)}\right)^3 \\ T^{(2)} = \alpha_0 + \alpha_{11}p_1^{(2)} + \alpha_{12}\left(p_1^{(2)}\right)^2 + \alpha_{13}\left(p_1^{(2)}\right)^3 + \\ \quad \ldots + \alpha_{N1}p_N^{(2)} + \alpha_{N2}\left(p_N^{(2)}\right)^2 + \alpha_{N3}\left(p_N^{(2)}\right)^3 \\ \quad \vdots \\ T^{(M)} = \alpha_0 + \alpha_{11}p_1^{(M)} + \alpha_{12}\left(p_1^{(M)}\right)^2 + \alpha_{13}\left(p_1^{(M)}\right)^3 + \\ \quad \ldots + \alpha_{N1}p_N^{(M)} + \alpha_{N2}\left(p_N^{(M)}\right)^2 + \alpha_{N3}\left(p_N^{(M)}\right)^3 \end{cases} \quad (1)$$

where $T^{(k)}$ is the value of total execution time of an application in k$^{th}$ experiment and $\left(p_1^{(k)}, p_2^{(k)}, \ldots, p_N^{(k)}\right)$ are

## Profiling and Modelling phase

1. For $i^{th}$ application in database ($\varphi_i$):
2.    For $j^{th}$ set of configuration parameters values $S_j = (M_j, R_j)$:
3.       Run $\varphi_i$ five times with $S_j$ for configuration parameters
4.       Assign average total execution time as the total execution time of the experiment
5.    End
6.    Form matrixes **A**, **T**, and **P** from Eqn.(2,3)
7.    Using Eqn.(6) calculate modelling coefficients of $\varphi_i$
8. End

(a)

## Prediction phase

1. Get values of configuration parameters $S_{user} = (M_{user}, R_{user})$
2. For $i^{th}$ application in database ($\varphi_i$):
3.    Upload $\varphi_i$'s individual Model
4.    Predict total execution time of $\varphi_i$ using Eqn.(5)
5. End

(b)

*Figure 2. Our proposed technique: (a) profiling and modeling, and (b) prediction algorithms*

the values of N chosen effective parameters for the same experiment, respectively. With matrix P as:

$$P = \begin{bmatrix} 1, p_1^{(1)}, (p_1^{(1)})^2, (p_1^{(1)})^3, \ldots, p_N^{(1)}, (p_N^{(1)})^2, (p_N^{(1)})^3 \\ 1, p_1^{(2)}, (p_1^{(2)})^2, (p_1^{(2)})^3, \ldots, p_N^{(2)}, (p_N^{(2)})^2, (p_N^{(2)})^3 \\ \vdots \\ 1, p_1^{(M)}, (p_1^{(M)})^2, (p_1^{(M)})^3, \ldots, p_N^{(M)}, (p_N^{(M)})^2, (p_N^{(M)})^3 \end{bmatrix} \quad (2)$$

Eqn.1 can be rewritten in matrix format as,

$$\underbrace{\begin{bmatrix} T^{(1)} \\ T^{(2)} \\ \vdots \\ T^{(M)} \end{bmatrix}}_{T} = P \begin{bmatrix} \alpha_0 \\ \alpha_{11} \\ \alpha_{12} \\ \alpha_{13} \\ \vdots \\ \alpha_{N1} \\ \alpha_{N2} \\ \alpha_{N3} \end{bmatrix}}_{A} \quad (3)$$

Using the above formulation, the approximation problem becomes to estimate the values of $\widehat{\alpha_0}, \widehat{\alpha_{11}}, \widehat{\alpha_{12}}, \widehat{\alpha_{13}} \ldots, \widehat{\alpha_{N1}}, \widehat{\alpha_{N2}}, \widehat{\alpha_{N3}}$ to optimize a cost function between the approximation and real values of total execution time. Then, an approximated total execution time $(\widehat{T^{(j)}})$ of the application for the $j^{th}$ experiment is predicted as:

$$\widehat{T^{(k)}} = \widehat{\alpha_0} + \widehat{\alpha_{11}} p_1^{(1)} + \widehat{\alpha_{12}} (p_1^{(1)})^2 + \widehat{\alpha_{13}} (p_1^{(1)})^3 + \cdots + \widehat{\alpha_{N1}} p_N^{(1)} + \widehat{\alpha_{N2}} (p_N^{(1)})^2 + \widehat{\alpha_{N3}} (p_N^{(1)})^3 \quad (4)$$

There are a variety of well-known mathematical methods in literature to calculate the variables $\widehat{\alpha_0}, \widehat{\alpha_{11}}, \widehat{\alpha_{12}}, \widehat{\alpha_{13}} \ldots, \widehat{\alpha_{N1}}, \widehat{\alpha_{N2}}, \widehat{\alpha_{N3}}$. One of these methods used widely in computer science and finance is Least Square Regression which calculates the parameters in Eqn.4 by minimizing the least square error as follows:

$$LSE = \sqrt{\sum_{i=1}^{M} (\widehat{T^{(i)}} - T^{(i)})^2}$$

The set of coefficients $\widehat{\alpha_0}, \widehat{\alpha_{11}}, \widehat{\alpha_{12}}, \widehat{\alpha_{13}} \ldots, \widehat{\alpha_{N1}}, \widehat{\alpha_{N2}}, \widehat{\alpha_{N3}}$ is the model that describes the relationship between the total execution time of the application regard to the configuration parameters. In another word, the approximate model between total execution time of the application and the configuration parameters is:

$$T = \mathcal{F}(p_1, p_2, \ldots, p_N) \\ = \widehat{\alpha_0} + \widehat{\alpha_{11}} p_1 + \widehat{\alpha_{12}} p_1^2 + \widehat{\alpha_{13}} p_1^3 \\ + \cdots + \widehat{\alpha_{N1}} p_N + \widehat{\alpha_{N2}} p_N^2 \\ + \widehat{\alpha_{N3}} p_N^3 \quad (5)$$

It can be mathematically proved that the least square error between real and approximated values is minimized when [30]

$$A = (P^T P)^{-1} P^T T \quad (6)$$

Figure 2-a is the algorithm for both profiling and modeling steps. An application ($\varphi_i$) is run for each set of the number of mappers and reducers and meanwhile total execution time of each experiment is extracted. Due to temporal changes in system, an experiment is repeated five time and then the average execution time is kept as the total execution time of that experiment. After finishing all experiments of the application, matrixes **A**, **T** and **P** are formed from Eqn. 2 and 3. Finally, the model coefficients of the application are calculated by Eqn6. This model later be used in prediction phase to predict the execution time of the same application for another set of the number of mappers and reducers.

### C. Prediction

Once a model has been created, it can then be applied to new experiments of the same application to calculate its estimated execution time for ranges of the number of Mappers and Reducers. It also should be considered that the model coefficients of an application may change from one application to another application and from one platform to another platform.

The prediction algorithm has been described in Figure 2-b. First the number of mappers and reducers for the same application are chosen by the user. Then the application model coefficients are used to estimate the execution time of the experiment by Eqn.3 and Eqn.5

## V. EXPERIMENTAL RESULTS

### A. Experimental setting

Two standard applications are used to evaluate the effectiveness of our method. Our method has been implemented and evaluated on a 4-node Map-Reduce cluster. Hadoop writes all files to the Hadoop Distributed File System (HDFS), and all services and daemons communicate over TCP sockets for inter-process communication. In our evaluation, the system runs Hadoop version 0.20.2 that is Apache implementation of Map-Reduce developed in Java [5]; the hardware specification of the nodes in our 4-node MapReduce platform is:

- Master/node-0 and node-1: Dell with one processor: 2.9GHz, 32-bit, 1GB memory, 30GB Disk, and 512KB cache.
- Node-2 and node-3: Dell with one processor: 2.5GHz, 32-bit, 512MB memory, 60GB Disk, and 254KB cache.

For each application in both profiling/modeling and prediction phases there are 20 sets of two configuration parameters values (number of Mappers and Reducers) where the number of Mappers and Reducers are chosen between 5 to 40 on 8GB of input data.

Our benchmark applications are WordCount, and Exim Mainlog parsing:

- *WordCount[33-34]:* This application reads data from a text file and counts the frequency of each word. Results are written in another text file; each line of the output file contains a word and the number of its occurrence, separated by a TAB. In running a WordCount application on Map-Reduce, each Mapper picks a line as input and breaks it into words $< key, value >$. Then it assigns a $< key, value >$ pair to each word as $< word, 1 >$. In the reduce stage, each Reducer counts the values of pairs with the same $key$ and returns occurrence frequency (the number occurrence) for each word,

- *Exim Mainlog parsing [35]*: Exim is a message transfer agent (MTA) for logging information of sent/received emails on Unix systems. This information that is saved in Exim Mainlog files usually results in producing extremely large files in mail servers. To organize such massive amount of information, a Map-Reduce application is used to parse the data – in an Exim Mainlog file – into individual transactions; each separated and arranged by a unique transaction ID.

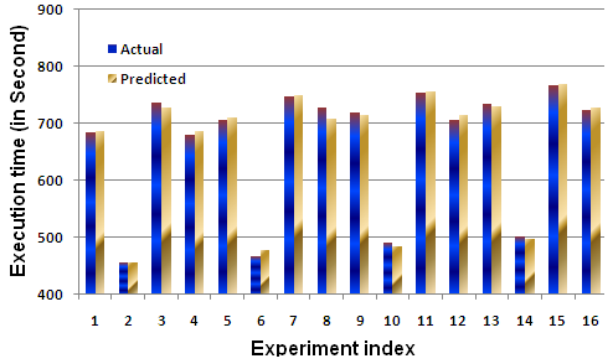
(a)

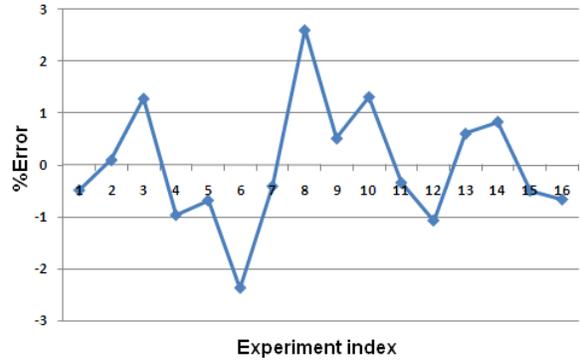
(b)

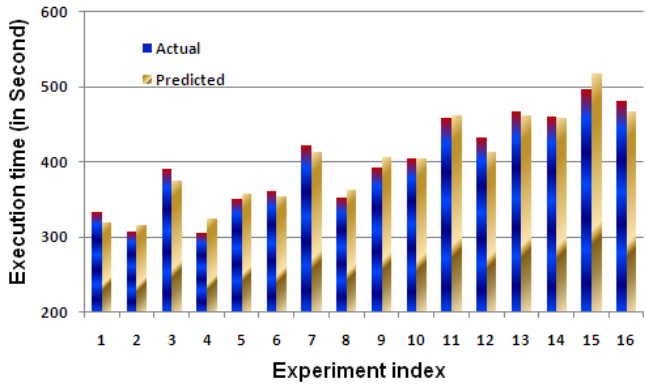
(c)

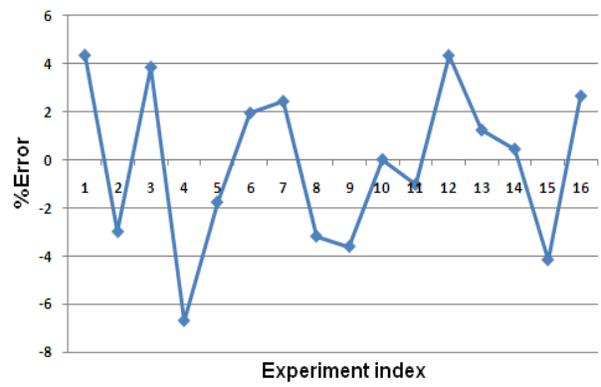
(d)

*Figure 3. the prediction accuracy and error between the actual and predicted total execution time for (a,b) WordCount, (c,d) Exim Mainlog parsing*

## B. Results

To test the accuracy of an application's model, we use it to predict execution time of several experiments of the application with random values of the number of Mappers and Reducers in the predefined range. We then run the experiments on a 4-node cluster and gather real execution time of the experiments to determine the prediction error.

For evaluation, the outcomes of these models on two standard MapReduce applications (WordCount written in Java, and Exim Mainlog parsing written in Python) are compared with their actual execution time. Figure 3 shows the prediction accuracies and prediction errors of these applications between actual execution times and their predicted execution times, while table 1 depicts the statistical mean and variance of prediction errors for the two applications.

We found that the average error between the actual execution times and their predicted counterparts is less than 5% for the tested applications. Some of the errors come from the model inaccuracy, but it can also be because of temporal changes in the system resulting in execution time increase for a short time, e.g. by background processes running during the execution of the applications. For example, in Hadoop, one of the main background processes comes from streaming when the Mapper and Reducer are written in a programming language other than Java. This is the reason about the big prediction error for Exim MainLog application as it was written in python.

As mentioned before, there is a strong dependency between a MapReduce application execution time and the number of Mappers and Reducers. Figure 4 shows the dependency between these two configuration parameters and execution time for both applications. As can be seen from this figure, the two applications behave differently for increasing the number of Mappers and Reducers. Generally, executing the same size of data for both applications results in different execution times so that, in most of time, WordCount has double execution time than Exim main log. In addition, although both applications show the minimum execution

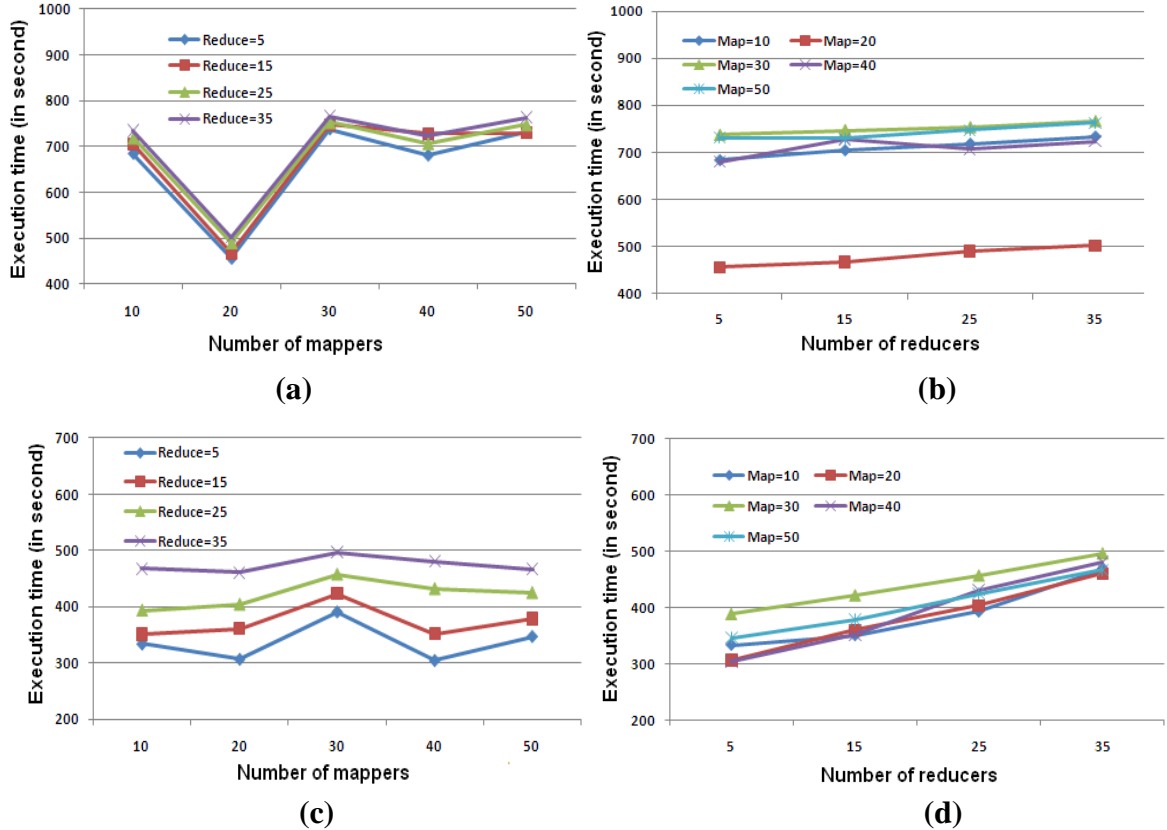

*Figure 4. the relation between total execution time and the number of Mappers and the number of Reducers for (a,b) WordCount, and (c,d) Exim Mainlog parsing*

time for 20 mappers and five reducers, WordCount shows more fluctuation than Exim for other number of mappers/reducers. The reason about why these number of mappers and reducers give the lowest execution time is not clear but it s too probable moving from one platform to another platform changes these numbers. Therefore, although the modeling is valid for applications on different platforms but the coefficients of the model may change by migration from one platform to another.

As reminder, although obtained models can successfully predict the value of total execution time required for a MapReduce application, they cannot give information about how application performance changes or how execution time varies during different MapReduce phases, such as Map and Reduce. Finally, our approach can help cloud customers and providers approximate the total execution time a MapReduce application needs in order to make scheduling jobs smarter.

## VI. CONCLUSION

This paper presents a new approach to model the correlation between two major MapReduce configuration parameters (number of Mappers and Reducers) and total execution time of applications when running on Map-Reduce clusters. After extracting the execution times of several experiments of an application with different values for the numbers of Mappers and Reducers, multivariate linear regression is used to model the relation between the extracted execution times and the used values for these two configuration parameters. Evaluation results for two standard applications on a 4-node MapReduce cluster show that our modeling technique can effectively predict the total execution time of these applications with the median prediction error of less than 5%.

## VII. ACKNOWLEDGMENT

The work reported in this paper is in part supported by National ICT Australia (NICTA). Professor A.Y. Zomaya's work is supported by an Australian Research Council Grant LP0884070.

TABLE 1. *The statistical mean and variance of prediction errors*

|  | **Mean (%)** | **Variance (%)** |
|---|---|---|
| **WordCount** | 0.9204 | 2.6013 |
| **Exim MainLog** | 2.7982 | 6.7008 |